\documentstyle[12pt]{article}
\begin{document}

\renewcommand{\thefootnote}{\fnsymbol{footnote}}

\begin{center}
{\large {\bf Relativistically Invariant Description of Fluctuations \\[0pt]
Against the Background of Classical Kink Solutions \\[0pt]
}} \vspace{1cm} V. D. Tsukanov \footnote{%
E-mail: tsukanov@kipt.kharkov.ua} \vspace{1cm}\\[0pt]
{\it Institute of Theoretical Physics}\\[0pt]
{\it National Science Center}\\[0pt]
{\it ''Kharkov Institute of Physics and Technology''}\\[0pt]
{\it 310108, Kharkov, Ukraine}\\[0pt]
\vspace{1.5cm}
\end{center}

\begin{quotation}
{\small {\rm A relativistically invariant scheme for the description of
excited states in a one-kink sector is formulated. The normal oscillations
of fluctuations against the background of a moving kink are determined. Zero
mode of these oscillations is excluded automatically due to the properties
of the integral equation describing normal oscillations. A Hamiltonian for
elementary excitations is obtained reflecting the relativistic nature of the
problem considered.}}
\end{quotation}

\renewcommand{\thefootnote}{\arabic{footnote}} \setcounter{footnote}0
\vspace{1cm}

\section{Introduction}

The idea to regard stable localized solutions of nonlinear equations as
dynamic objects has been under intense study in a number of papers published
in the middle of seventies (for references see \cite{R}). In the
one-dimensional case a coordinate $X$ of the kink solution $u_c(x-X)\ $of
the Klein-Gordon nonlinear equation has been regarded as a dynamic variable
and the initial field has been written in the form \cite{CL,T,J}
\begin{equation}
\Phi (x)=u_c(x-X)+\chi (x-X)\,.
\end{equation}
The coupling condition due to fluctuation
\begin{equation}
\left\langle \chi u_c^{\prime }\right\rangle \equiv \int dx\chi
(x)u_c^{\prime }(x)=0
\end{equation}
conserved the number of independent variables in the configurational space
and removed the zero mode $u_c^{\prime }$ from the spectrum of excitations.
At the same time it was noted that decomposition of the field into a
solitonic and fluctuating parts (1) is relativistically non-invariant and
first of all is adapted to studying the system properties in the
center-of-mass frame. In this case the Ansatz $u_c^{\prime }(x)$ describes
the classic field configuration in the ground state exactly and the variable
$\chi (x)$ describes small perturbations related to the presence of field
fluctuations as well as to kink trembling. Thus, e.g., the quantum
corrections to the soliton mass \cite{J} have been determined and the system
spectrum in the center-of-mass frame \cite{R} has been studied. Besides, as
the relativistic invariance of the Hamiltonian does not depend on the
concrete choice of variables, there has always existed a possibility to
restore the relativistic form of the heavy particle energy in one way or
other. In Ref. \cite{GJeS} this problem has been solved via summation of
tree diagrams of the perturbation theory and Ref. \cite{O} has found the
corresponding special solutions of the equations of motion for the
fluctuations. At the following on the ground of the expansions of type (1)
the studies of meson-soliton scattering in one-dimensional model have been
attempted \cite{O,O2,HSaU,UHSa}. Similar studies have also been made in
Skyrme model \cite{SaOtY,LiOk,HSaU2,HSaU3}. The total momentum in these
problems does not vanish and therefore it is unjustified to treat the field $%
\chi (x)$ as a purely fluctuational one. This treatment fails to take into
account the translational kink motions and contradicts the relativistic
nature of the theory.

Concerning a touched problem we note, that by virtue of a relativism, the
investigation of a system in the center-of-mass reference frame allows, in
principle, to restore a character of its evolution in any frame of
reference. However, such passage should be adjust with used perturbation
theory and its details are not quite clear. Therefore, it is desirable, from
the very beginning to present the scheme in the relativistic invariant form.
For this purpose let us note some properties of one kink states. First of
all, note that one cannot introduce a relativistically invariant vacuum
state in a topological sector in contrast to the conventional field theory.
The self-similar solutions of equations of motion in the form of moving
kinks realize only local minima of the Hamiltonian that can be parametrized
with the total momentum of the system. From the relativistic point of view
these minima are equivalent, degenerate with respect to spatial translations
and absolutely stable with respect to arbitrary fluctuations of field
variables due to the conservation of the total momentum of the system. Thus,
when one studies the topological sector of the theory from the viewpoint of
an arbitrary reference frame, one should find a local minimum relating to
the respective value of the total momentum of the system instead of defining
the relativistically invariant vacuum state. To study the corresponding
variational problem one should introduce the cyclic coordinate $X$ as a
collective variable with the help of the substitution
\begin{equation}
\Phi (x)=\varphi (x-X)
\end{equation}
imposing on the function $\varphi (x)$ certain limitations providing for the
non-degenerate nature of the transformation to new variables. Developing the
canonical procedure on the ground of the variables $X$, $\varphi (x)$ and
minimizing the Hamiltonian with respect to variations of the field $\varphi
(x)$ and the momentum conjugate to it yield a usual root at the extremum
point defining the energy of a relativistic particle. The momentum of this
particle coincides with the total momentum of the system and its mass
corresponds to the kink's mass. Fluctuations of field variables against the
local minimum are defined in a relativistically invariant way and the study
of their spectrum permits one to present the Hamiltonian in the form of a
series in powers of the amplitudes of corresponding normal oscillations. The
terms of this series depend on the total momentum of the system that is not
an observable quantity from the physical viewpoint. Splitting the total
momentum of the system into the kink and field parts one can distinguish the
individual dynamic variables of the kink and the mesons. As the field
momentum is small due to small fluctuations, one should again expand the
Hamiltonian of the system in powers of the meson field amplitudes.
Performing these procedures, one can obtain the Hamiltonian adapted for
calculating the meson-soliton scattering effects. The paper shows that the
kinetic term of this Hamiltonian describes the multitude of relativistic
particles including a heavy particle - kink and a set of light particles
corresponding to the states of the background scattering of mesons by the
kink. Such a structure of the system spectrum in the presence of a kink has
been noted in Ref. \cite{GoJa}. A detailed treatment of the outlined scheme
is divided into three Sections and an Appendix. In the next Section a local
minimum is determined and the part of the Hamiltonian bilinear in
fluctuations is calculated. Section 3 deals with the normal oscillations of
the system. Section 4 introduces amplitudes of the meson field and
calculates the kinetic part of the Hamiltonian. In the Appendix the
indefinite dot product used for the field quantization is derived.

\section{Determining a local minimum and the Hamiltonian of fluctuations}

Consider a nonlinear scalar system described by the Lagrangian.
\[
L=\int dx\left\{ \frac 12\left( \dot{\Phi}^2(x)-\Phi ^{\,\prime
\,2}(x)\right) -U\left( \Phi (x)\right) \right\} \,.
\]
Kink solutions to this system realize a multitude of local minima of the
Hamiltonian degenerated with respect to spatial translations. Taking into
account the dominant role of degeneration parameters of vacuum solutions in
the formation of a low-energy spectrum, let us introduce a cyclic coordinate
$X$ as a collective variable with the help of substitution (3) limiting the
admissible variations of the field $\varphi (x)$ by the condition
\begin{equation}
\left\langle \overline{\Phi }^{\ \prime }\delta \varphi \right\rangle =0\,.
\end{equation}
$\overline{\Phi }(x)\ $is the time-independent field Ansatz whose form will
be determined later. Taking into account that velocity variations $\delta
\dot{\varphi}(x)$ also satisfy the coupling condition $\left\langle
\overline{\Phi }^{\ \prime }\delta \dot{\varphi}\right\rangle =0\,$, we find
the canonical momenta conjugate to new variables
\begin{equation}
P=\partial L/\partial \dot{X}=\left\langle (\dot{X}\varphi ^{\,\prime }-\dot{%
\varphi})\varphi ^{\,\prime }\right\rangle \,,\qquad \pi (x)=\delta L/\delta
\dot{\varphi}(x)=(1-{\cal P)}\left( \dot{\varphi}(x)-\dot{X}\varphi
^{\,\prime }(x)\right) \,,
\end{equation}
where
\begin{equation}
{\cal P}(x,\ x^{\,\prime })=E^{-1}\overline{\Phi }^{\ \prime }(x)\overline{%
\Phi }^{\ \prime }(x^{\,\prime })\,,\qquad E\equiv \left\langle \overline{%
\Phi }^{\ \prime \,2}\right\rangle \,,\qquad {\cal P}^2={\cal P}
\end{equation}
is the projection operator. Solving relation (5) with respect to the
velocities $\dot{X}$, $\dot{\varphi}(x)\ $we find the Hamiltonian of the
system in new variables
\begin{equation}
H=\dot{X}P+\left\langle \dot{\varphi}\,\,\pi \right\rangle -L=\frac
12\left\{ \left\langle \pi ^2+\varphi ^{\,\,\prime \,2}+2U(\varphi
)\right\rangle +\frac{\left( \left\langle \varphi ^{\,\prime }\pi
\right\rangle +P\right) ^2}{\left\langle \varphi ^{\,\prime }{\cal P}\varphi
^{\,\prime }\right\rangle }\right\} \,.
\end{equation}
The equations for the extremals $\overline{\varphi }(x),\ \overline{\pi }(x)$
\begin{equation}
\delta H/\delta \pi =0\,,\qquad \delta H/\delta \varphi =0
\end{equation}
determine the local minima of the Hamiltonian parametrized by the total
momentum of the system $P\,$. As, according to (5), the variations of the
momentum satisfy the condition ${\cal P}\delta \pi =0\,$, then the first of
equations (8) assumes the form
\begin{equation}
(1-{\cal P})\left\{ \overline{\pi }+\overline{\varphi }^{\,\prime }\frac{%
P+\left\langle \varphi ^{\,\prime }\pi \right\rangle }{\left\langle \varphi
^{\,\prime }{\cal P}\varphi ^{\,\prime }\right\rangle }\right\} =0\,.
\end{equation}
Let us now fix the gauge Ansatz $\overline{\Phi }(x)$ identifying it with
the extremal of the Hamiltonian: $\overline{\Phi }(x)\equiv \overline{%
\varphi }(x)$. Then from equation (9) it follows that $\overline{\pi }(x)=0$
and the second of equations (8) assumes the form 
\[
-\left( 1-\frac{P^2}{E^2}\right) \overline{\Phi }^{\,\prime \prime
}+U^{\prime }(\overline{\Phi })=0\,. 
\]
That is, 
\[
\overline{\Phi }(x)=u_c\left( \frac E{\sqrt{E^2-P^2}}\,x\right) \,, 
\]
where $u_c(x)$ is the static kink solution. Substituting the solutions $%
\overline{\pi }(x)=0$, $\overline{\varphi }(x)=\overline{\Phi }(x)\ $into
the definition of the parameter $E\ $(6) and in formula (7) we prove that
the local minimum of the Hamiltonian coincides with the energy of a
relativistic particle 
\begin{equation}
H_0=E=\sqrt{P^2+M^2\,},
\end{equation}
$M\equiv <u_c^{\prime 2}>$ is the kink's mass. Fluctuations of canonical
variables $\varphi (x)-\overline{\varphi }(x)$, $\pi \left( x\right) \ $are
obviously defined in a relativistically invariant way. Keeping for them the
initial notation, let us write the Hamiltonian bilinear in these
fluctuations 
\begin{equation}
H_2=\frac 12\left\langle \pi ^2+\varphi ^{\,\prime \,2}+2\frac PE\varphi
^{\,\prime }\pi +U^{\prime \prime }(\overline{\Phi })\varphi ^2+3\frac{P^2}{%
E^2}\varphi ^{\,\prime }{\cal P}\varphi ^{\,\prime }\right\rangle \,.
\end{equation}
According to (4), (5) Poisson's brackets of the variables $\varphi (x),\ \pi
(x),\ P$ have the form 
\begin{equation}
\left\{ \varphi (x),\pi (x^{\prime })\right\} =(1-{\cal P})(x,x^{\prime
})\,,\qquad \left\{ P,\varphi \right\} =\left\{ P,\pi \right\} =0\,,\qquad
\left\{ \varphi ,\varphi \right\} =\left\{ \pi ,\pi \right\} =0\,.
\end{equation}

\section{Normal oscillations}

The Hamiltonian $H_2$ describes the small oscillations of the system in the
vicinity of a local minimum. Using (11), (12) we will write the
corresponding equations of motion for the variables $\varphi (x),$ $\pi
\left( x\right) \ $as 
\begin{eqnarray*}
\dot{\pi}(x) &\equiv &-\frac{\delta H_2}{\delta \varphi (x)}=(1-{\cal P}%
)\left( \frac PE\frac \partial {\partial x}\pi \left( x\right) -{\cal L}%
_p\varphi (x)+3\frac{P^2}{E^2}\frac \partial {\partial x}{\cal P}\frac
\partial {\partial x}\varphi (x)\right) \,, \\
\dot{\varphi}(x) &\equiv &-\frac{\delta H_2}{\delta \pi (x)}=(1-{\cal P}%
)\left( \pi (x)+\frac PE\frac \partial {\partial x}\varphi \left( x\right)
\right) ,\qquad {\cal L}_P\equiv -\frac{\partial ^2}{\partial x^2}+U^{\prime
\prime }(\overline{\Phi })\,.
\end{eqnarray*}
Rewriting the second equation in the form 
\begin{equation}
\pi (x)=(1-{\cal P})D\varphi (x)\,,\qquad D\equiv \frac \partial {\partial
t}-\frac PE\frac \partial {\partial x}\,,
\end{equation}
we find the equation for the field fluctuations $\varphi (x)$%
\begin{equation}
\left( {\cal L}_P+D^2-4\frac{P^2}{E^2}\frac \partial {\partial x}{\cal P}%
\frac \partial {\partial x}\right) \varphi (x)=0\,.
\end{equation}
To solve this equation, we define a new variable $f(z)$ according to the
formula 
\begin{equation}
\varphi (x,t)=e^{-i\frac \omega E(Mt-Pz)}f(z)\,,
\end{equation}
$z\equiv (E/M)x\ $is the scale transformation of the coordinate. Using the
explicit form of the operators ${\cal L}_p,\ D\ $and differentiating, we
find 
\[
\left( {\cal L}_P+D^2\right) e^{-i\frac \omega E(Mt-Pz)}f(z)=e^{-i\frac
\omega E(Mt-Pz)}({\cal L}-\omega ^2)f(z)\,,\qquad {\cal L\equiv }-\frac{%
\partial ^2}{\partial z^2}+U^{\prime \prime }(\overline{\Phi })\,. 
\]
Then, taking into account that the function $\overline{\Phi }^{\,\prime
}(x)\ $is zero mode of the operator ${\cal L\ }$and using the coupling
condition (4) written in the form 
\begin{equation}
{\cal P}e^{i\,\omega \frac PEz}f(z)=0\,,
\end{equation}
we obtain from (14) the equation for the function $f(z)$ 
\begin{equation}
({\cal L}-\omega ^2)f(z)=-F(\omega ,z)\int dz^{\prime }F^{*}(\omega
,z^{\prime })f(z^{\prime })\equiv -F(\omega ,z)Q(\omega )\,,
\end{equation}
where 
\[
F(\omega ,z)\equiv e^{-i\,\omega \,\frac PEz}\left( \frac ME\omega -2i\frac
PM\frac \partial {\partial z}\right) \psi _0(z) 
\]
and 
\[
\psi _0(z)=\frac 1{\sqrt{M}}u_c^{\prime }(z)=\frac{\sqrt{M}}E\overline{\Phi }%
^{\,\prime }(x),\qquad \int dz\psi _0^{\,2}(z)=1 
\]
is the zero mode of the operator normalized to unity in the variables $z$.
Formally one may regard equation (17) as a linear equation with a right-hand
side. The solution of a homogeneous problem 
\begin{equation}
{\cal L}\psi _\alpha (z)=\omega _\alpha ^2\psi _\alpha (z),\qquad \alpha
\equiv \{i,q\}\,,
\end{equation}
determines the frequencies of normal oscillations of the system at rest. To
solve the equation (17) at the points of the spectrum $\omega =\omega
_{\alpha \ }$of the operator ${\cal L\ }$the eigenstates $\psi _\alpha \ $%
must be orthogonal to the right-hand side of this equation 
\begin{equation}
\int dz\psi _\alpha (z)F(\omega _\alpha ,z)=0\,.
\end{equation}
At the same time the corresponding solutions of equation (17) will be
written in the form 
\begin{equation}
f_\alpha (z)=\psi _\alpha (z)-({\cal L}-\omega _\alpha ^2)^{-1}F(\omega
_\alpha ,z)Q(\omega _\alpha )\,.
\end{equation}
Inserting this solution in the definition (17) for the constants $Q(\omega
_\alpha )\ $and using (19) yield the equation 
\begin{equation}
\left( 1+\left\langle F^{*}(\omega )({\cal L}-\omega _\alpha
^2)^{-1}F(\omega )\right\rangle _z\right) Q(\omega _\alpha )=0\,,\qquad
<...>_z\equiv \int dz...
\end{equation}
For the zero mode $\omega _\alpha =0\ $equation (21) possesses only zero
solution $Q(0)=0\ $because the operator ${\cal L\ }$is positively defined.
As the function $f_0\ $(20) coincides with the zero mode $\psi _0$, the
gauge condition (16) is not met for it and, consequently, one should exclude
the mode $f_0\ $from the fluctuation spectrum. One should note in this
connection that the form of the equation for the function $f(z)\ $(17) is
not related to the gauge choice for the function $\varphi (x)\ $(3) \cite{Ts}%
. Only the concrete values of the constants $Q(\omega _\alpha )\ $depend on
this choice. The constant $Q(0)\ $vanishes with any gauge. Therefore the
function $f_0\ $(20) coinciding with zero mode of the operator ${\cal L\ }$%
(18) should be excluded from the spectrum of excitations because a fixed
function cannot satisfy an arbitrary gauge. For other modes ($\omega _\alpha
\neq 0$) the parameters $Q(\omega _\alpha )$, generally speaking, must be
different from zero and must ensure the fulfilment of arbitrary gauge
conditions. This means that the following equality must hold 
\begin{equation}
1+\left\langle F^{*}(\omega )({\cal L}-\omega _\alpha ^2)^{-1}F(\omega
)\right\rangle _z=0\,,\qquad \omega _\alpha \neq 0\,.
\end{equation}
Performing the integration by parts,$\ $one may show that the eigenfunctions
of the operator ${\cal L\ }$(18) satisfy the identity 
\[
2i\upsilon \int dze^{-i\upsilon \,z}\psi _\alpha (z)\psi _0^{\,\prime
}(z)=(\omega _\alpha ^2-\upsilon ^2)\int dze^{-i\upsilon \,z}\psi _\alpha
(z)\psi _0(z) 
\]
that establishes the link between the integrals in both parts of this
equality for arbitrary $\upsilon \,$. Using this link and the relation $%
E^2=M^2+P^2$, one proves easily the validity of the solubility condition
(19). The same identity together with the completeness condition 
\[
\sum_\alpha \psi _\alpha (z)\psi _\alpha ^{*}(z^{\prime })=\delta
(z-z^{\prime })\,,\qquad \sum_\alpha \equiv \int dq+\sum_i 
\]
permits one to prove formula (22) for arbitrary $\omega \neq 0\,$.
Similarly, one can get rid of the resolvent $({\cal L}-\omega _\alpha
^2)^{-1}\ $in (20) and write the solution of equation (17) in the form 
\[
f_\alpha (z)=\psi _\alpha (z)+\frac E{M\omega _\alpha }e^{-i\,\omega _\alpha
\frac PEz}\psi _0(z)Q(\omega _\alpha ). 
\]
Determining the constants $Q(\omega _\alpha )\ $from the gauge condition
(16), one may write this solution in the form 
\begin{equation}
f_\alpha (z)=(1-{\cal P}_\alpha )\psi _\alpha (z),
\end{equation}
where 
\[
{\cal P}_\alpha (z,z^{\prime })=e^{-i\,\omega _\alpha \frac PEz}\psi
_0(z)\psi _0(z^{\prime })e^{i\,\omega _\alpha \frac PEz^{\prime }\,},\qquad 
{\cal P}_\alpha ^2={\cal P}_\alpha 
\]
is the projection operator. Obviously, the solution of equation (14) splits
into components with positive and negative frequencies. If one denotes with $%
\omega _i^2\neq 0,\ i=1,...,N;\ \omega _q^2=q^2+m^2\ $the eigenvalues of the
operator ${\cal L}\ $belonging to the states of the discrete and continuous
spectra, where $q\ $is the wave vector of the states from the continuous
spectrum, $m^2$ being its boundary, then, according to (15), (23), one may
write all multitude of solutions of equation (14) determining the normal
oscillations in the vicinity of the local minimum in the form 
\begin{equation}
\varphi _\alpha ^{(+)}(x,t)=\frac 1{\sqrt{2\omega _\alpha }}exp\left( -i%
\frac{\omega _\alpha }E(Mt-Pz)\right) (1-{\cal P}_\alpha )\psi _\alpha
(z),\quad \varphi _\alpha ^{(-)}=\varphi _\alpha ^{(+)*},\quad \alpha =i,\ q.
\end{equation}
Here and in what follows $\omega _\alpha \ $are positive quantities.
Functions $\psi _\alpha \ $in (24) are normalized conventionally in $L_2$%
\begin{equation}
\int dz\psi _q^{*}(z)\psi _{q^{\,\prime }}(z)=\delta (q-q^{\,\prime }),\
\qquad \int dz\psi _i(z)\psi _j(z)=\delta _{ij}\,,
\end{equation}
and the orthogonality properties of the functions $\varphi _\alpha ^{(\pm
)}(x,t)\ $and their normalization with respect to the indefinite dot product
(A4) are given in the Appendix.

\section{Elementary excitations}

Expanding the field $\varphi (x,t)\ $over the functions orthonormalized in
the sense of the dot product (A4) 
\begin{equation}
\varphi (x,t)=\sum_\alpha \left( c_\alpha \varphi _\alpha
^{(+)}(x,t)+c_\alpha ^{*}\varphi _\alpha ^{(-)}(x,t)\right)
\end{equation}
and using (26), (13) let us express the amplitudes $c_\alpha ,\ c_\alpha
^{*}\ $through the canonical variables $\varphi (x),\ \pi (x)$ 
\begin{equation}
c_\alpha =(\varphi _\alpha ^{(+)},\varphi )=i\left\langle \varphi _\alpha
^{(-)}\pi -(D\varphi _\alpha ^{(-)})\varphi \right\rangle \,,\qquad c_\alpha
^{*}=-(\varphi _\alpha ^{(-)},\varphi )=-i\left\langle \varphi _\alpha
^{(+)}\pi -(D\varphi _\alpha ^{(+)})\varphi \right\rangle \,.
\end{equation}
With the help of formulas (12), (A6), (A7), (27), one may calculate the
Poisson's brackets of a new set of variables $P,\ c_\alpha ,\ c_\alpha
^{*}\, $. 
\begin{equation}
\{P,c_\alpha \}=\{P,c_\alpha ^{*}\}=0,\qquad \{c_\alpha ,c_{\alpha ^{\prime
}}\}=\{c_\alpha ^{*},c_{\alpha ^{\prime }}^{*}\}=0,\qquad \{c_\alpha
,c_{\alpha ^{\prime }}^{*}\}=-i\delta _{\alpha \alpha ^{\prime }}\,.
\end{equation}
To write (11) in terms of the amplitudes, let us present formally the
nonlocal Hamiltonian bilinear in the field variables $\varphi (x),\ \pi (x)\ 
$in the form 
\begin{equation}
H_2=(1/2)\left\langle \varphi V\varphi +2\varphi K\pi +\pi S\pi
\right\rangle \,,
\end{equation}
where in the general case $V,\ K,\ S\ $are some integro-differential
operators. Canonical equations of motion can be written in the form 
\[
V\varphi =-\dot{\pi}-K\pi \,,\qquad \pi S=\dot{\varphi}-\varphi K\,. 
\]
Inserting these expressions into the first and third terms of (29),
respectively, we find 
\begin{equation}
H_2=(1/2)\left\langle \dot{\varphi}\,\pi -\varphi \,\dot{\pi}\right\rangle
\,.
\end{equation}
In this Hamiltonian the role of dynamic variables is played by the
integration constants of the equations of motion. Substituting expressions
(26), (13) in (30) with the account of the temporal dependence (15) and
using formulas (A4-A7), let us write the Hamiltonian $H_2\ $in terms of the
amplitudes\footnote{%
One can restore the spectrum of this Hamiltonian obtained under the
condition that the\ amplitude are small from the spectrum of the system at
rest with the help of formula $E^{(n)}(P)=\sqrt{P^2+E^{(n)}(0)}$ \cite
{GJeS,R}} 
\begin{equation}
H_2=\frac ME\left( \int dq\omega _{\,q}c_q^{*}c_q+\sum_i\omega
_ic_i^{*}c_i\right) ,\qquad \omega _{\,q}=\sqrt{q^2+m^2}\,.
\end{equation}

Using formulas (24), (26) one can also write the next terms of the expansion
of the Hamiltonian in terms of the amplitudes of normal oscillations $%
c_\alpha \,,\ c_\alpha ^{*}\,$. The Hamiltonian thus obtained describes the
excitations of the system against the local topologic minimum with the given
value of the total momentum of the system. These excitations possess a
hybrid nature taking into account meson field oscillations and kink's
trembling simultaneously. To study the individual movements of the kink and
the mesons and to describe the interaction between them one needs to split
the total momentum of the system into the purely kink part $p$ and one
related to field fluctuations $P_f\,$: $P=p+P_f\,$. After that one should
re-expand the obtained Hamiltonian using the small value of the field
momentum $P_f$. To determine the kinetic part of the Hamiltonian
restructured in this way, it is sufficient to use the momentum $P_f$ in the
leading approximation. One can establish the momentum of fluctuations as
follows. The wave vector $q$ parametrizing the eigenfunctions and
eigenvalues of the continuous spectrum of problem (18) in the leading (zero)
approximation with respect to the reflection coefficient is the momentum of
the corresponding state in the center-of-mass reference frame. In the
reference frame relative to which the system moves with the velocity $%
V=P/E\, $, the momentum of this state $\overline{q}$ and its energy $\omega
_{\overline{q}}$ in accord with the Lorentz transformations will amount to 
\begin{equation}
\overline{q}=\frac{q+V\omega _q}{\sqrt{1-V^2}}\,,\qquad \omega _{\overline{q}%
}=\frac{\omega _q+Vq}{\sqrt{1-V^2}}\,.
\end{equation}
Therefore, multiplying the momentum $\bar{q}$ by the number of ''particles''
in this state $c_q^{*}c_q$, and performing the integration over the
parameter $q$, we find the momentum of the field in the reference frame
relative to which the system moves with the velocity $V=P/E$%
\[
P_f=\int dq\overline{q}c_q^{*}c_q\,. 
\]
As $dq/d\overline{q}=\omega _q/\omega _{\overline{q}}\,$, then passing to
the integration over $\overline{q}\ $in this formula and introducing the
meson amplitudes 
\begin{equation}
a_{\overline{q}}=\sqrt{\frac{\omega _q}{\omega _{\overline{q}}}}c_q\,,\qquad
a_{\overline{q}}^{*}=\sqrt{\frac{\omega _q}{\omega _{\overline{q}}}}%
c_q^{*}\,,
\end{equation}
we write the momentum of fluctuations in the usual form 
\begin{equation}
P_f=\int dqqa_q^{*}a_q\,,
\end{equation}
Note that the expression for the momentum of fluctuations (34) may be
obtained by separating the contribution proportional to the volume of the
system from the integral $-<\pi \varphi ^{\,\prime }>\ $and neglecting the
reflection effects. Using (28), (33) one can prove that the Poisson's
brackets of the variables $p,\ a_q,\ a_q^{*}\ $have the form 
\begin{eqnarray*}
\{a_q,a_{q^{\,\prime }}\} &=&\{a_q^{*},a_{q^{\,\prime }}^{*}\}=0\,,\qquad
\{a_q,a_{q^{\,\prime }}^{*}\}=-i\delta (q-q^{\,\prime })\,, \\
\{p,a_q\} &=&-iqa_q\,,\qquad \{p,a_q^{*}\}=iqa_q^{*}\,.
\end{eqnarray*}
Taking into account formulas (10), (31), (33) and keeping the linear term in 
$P_f$ in Hamiltonian $H_0$, we find 
\begin{eqnarray*}
H_0 &=&\sqrt{p^2+M^2}+V\int d\overline{q}\,\overline{q}\,a_{\overline{q}%
}^{*}\,a_{\overline{q}}+O(a^4)\,, \\
H_2 &=&\sqrt{1-V^2}\int d\overline{q}\,\omega _qa_{\overline{q}}^{*}\,a_{%
\overline{q}}+\left( M/\sqrt{p^2+M^2}\right) \sum_i\omega
_ic_i^{*}c_i+O\left( (ca)^2\right) \,.
\end{eqnarray*}
Hence summing main contributions in amplitudes and using transformations
(32) we find kinetic term of full Hamiltonian. 
\[
{\cal H}_0=\sqrt{p^2+M^2}+\int dq\omega _q\,a_q^{*}\,a_q+\left( M/\sqrt{%
p^2+M^2}\right) \sum_i\omega _ic_i^{*}c_i\,. 
\]
This Hamiltonian reflects the natural pattern of the spectrum in the
one-kink spectrum sector noted in work \cite{GoJa} and describes a free
relativistic particle-kink, a set of light particles related to the states
of background scattering of mesons on a kink as well as excitations of the
internal degrees of freedom.

\section{Conclusion}

The paper formulates the relativistically invariant procedure for the
description of excited states in a one-kink sector. The main part of the
paper deals with studying the dynamics of the system with a fixed total
momentum. It includes the determination of the local minimum of the
Hamiltonian formally corresponding to a free moving kink as well as studies
of fluctuations in the vicinity of this minimum. A relativistic form of the
corresponding normal oscillations is established. Zero mode of these
oscillations is excluded automatically due to the properties of the integral
equation describing normal oscillations. In this connection it should be
noted that the role of the gauge condition imposed on the function $\varphi
(x)\ $(3) reduces only to the non-contradictory introduction of new
variables. It is also evident that the choice of the gauge affects the form
of equations of motion. The gauge condition used in the work is defined by
the relativis-tically invariant way. It leads to zero canonical momentum of
fluctuations in the extremum point and essentially to simplify the equations
of motion. Keeping in mind that with the fixed momentum of the system the
fluctuations have a hybrid pattern describing the oscillations of the meson
field as well as the kink's trembling, the momentum of the system is split
into field and kink parts in the final part of the paper. This permits one
to attribute to the kink the sense of a full-scale dynamic object and to
write the kinetic term of the Hamiltonian in terms of natural elementary
excitations characteristic for a relativistic theory.\newline

%\renewcommand{sectionnumber}{2}

%\section
{\Large {\bf {Appendix}} } \newline

Let us use the identity 
$$
\left\langle \varsigma ^{*}\left( {\cal L}_P+D^2-4\frac{P^2}{E^2}\frac
\partial {\partial x}{\cal P}\frac \partial {\partial x}\right) \varphi
\right\rangle \,=0\,, \eqno{A1} 
$$
where $\varsigma \,,\ \varphi \ $are the solutions of equation (14) to
introduce an indefinite dot product applied for the field quantization.
Performing the integration by parts, we obtain 
\[
\frac \partial {\partial t}\left\langle \varsigma ^{*}(D\varphi
)-(D\varsigma ^{*})\varphi \right\rangle +\frac PE\left. \left( (D\varsigma
^{*})\varphi -\varsigma ^{*}(D\varphi )\right) \right| _{-\infty }^{+\infty
}+\left. \left( \varsigma ^{*\,\prime }\varphi -\varsigma ^{*}\varphi
^{\,\prime }\right) \right| _{-\infty }^{+\infty }=0\,. 
\]
Introducing according to (15), the substitution 
\[
\varphi (x,t)=exp\left( -\frac PEz\frac \partial {\partial t}\right)
h(z,t)\,,\qquad \varsigma (x,t)=exp\left( -\frac PEz\frac \partial {\partial
t}\right) g(z,t) 
\]
and using it for the transformation of the terms located outside the
integral sign, we get 
$$
\frac \partial {\partial t}\left\langle \varsigma ^{*}(D\varphi
)-(D\varsigma ^{*})\varphi \right\rangle +\frac MEexp\left( -\frac PEz\frac
\partial {\partial t}\right) \left. \left( \frac{\partial g^{*}}{\partial z}%
h-g^{*}\frac{\partial h}{\partial z}\right) \right| _{-\infty }^{+\infty
}=0\,. \eqno{A2} 
$$
The term outside the integral here is not vanishing only for functions $g,\
h\ $belonging to the continuous spectrum and possessing the same sign of the
frequency. As for the concrete modes with the wave vectors $q^{\,\prime
}\simeq \pm q$%
\[
\omega _{q^{\,\prime }}-\omega _q\simeq \pm \frac q{\omega _q}(q^{\,\prime
}\mp q)\,, 
\]
where $\omega _q\ $is the positive value of the root $\sqrt{q^2+m^2}$, then
taking into account the temporal dependence (15) one can change the
exponential factor in (A2) with unity. Indeed, after differentiating with
respect to $t$, the exponent of this factor will be equal to 
\[
\frac PE(\omega _q-\omega _{q^{\,\prime }})z\simeq \pm \frac PE\frac
q{\omega _q}(q^{\,\prime }\mp q)z\,. 
\]
As $\left| (P/E)(q/\omega _q)\right| <1\ $then due to formula 
$$
q^{-1}exp(\pm \,iqx)\left.\right |_{x\rightarrow \infty }\rightarrow\pm
\,i\pi \delta (q) \eqno{A3} 
$$
this exponential factor will not influence the values of the limits in (A2)
that are determined by the phases $(q^{\,\prime }\mp q)z\ $of the second
factor in this term. Thus, if one defines the indefinite dot product with
the formula\footnote{%
This definition is similar to the dot product $(\varsigma ,\varphi )\equiv
-i\int dx(\dot{\varsigma}^{*}\varphi -\varsigma ^{*}\dot{\varphi})$used
while quantizing the Klein-Gordon free field \cite{BD}} 
$$
(\varsigma ,\varphi )\equiv -i\left\langle (D\varsigma ^{*})\varphi
-\varsigma ^{*}(D\varphi )\right\rangle =-(\varphi ,\varsigma )^{*}\,, %
\eqno{A4} 
$$
then for the functions satisfying equation (14) the identity (A1) may be
written in the form 
$$
\frac \partial {\partial t}(\varsigma ,\varphi )=i\frac ME\left. \left(
g_0^{*}\frac{\partial h_0}{\partial z}-\frac{\partial g_0^{*}}{\partial z}%
h_0\right) \right| _{-\infty }^{+\infty }\,. \eqno{A5} 
$$
Here $g_0,\ h_0\ $are the asymptotics of the functions $g,\ h\ $at $\left|
x\right| \rightarrow \infty .$ Without accounting for the phase shifts they
are defined by the equation $(\partial ^2/\partial t^2-\partial ^2/\partial
z^2+m^2)g_0=0$. Hence with the account of the temporal dependence and using
formula (A3)\ there follows that 
$$
(\varphi ^{(-)},\varphi ^{(+)})=0\,,\qquad (\varphi _i,\varphi _q)=0\,. %
\eqno{A6} 
$$
Besides, the functions (24) are obey the normalisation conditions 
$$
(\varphi _{\,q}^{(\pm )},\varphi _{q^{\,\prime }}^{(\pm )})=\pm \,\delta
(q-q^{\,\prime })\,,\qquad (\varphi _{\,i}^{(\pm )},\varphi _{\,j}^{(\pm
)})=\pm \,\delta _{ij}\,. \eqno{A7} 
$$
The first of formulas (A7) is obvious actually due to (A5) and the
assumption about the normalization of the functions $\psi _q\ $of the
continuous spectrum in $L_2\ $(25). One can prove the validity of the second
one using formula (24) in the direct calculation of the integral (A4) and
including the normalization of the functions $\psi _i\ $of the discrete
spectrum in $L_2\ $(25).


\begin{thebibliography}{99}
\bibitem{R}  R. Rajaraman, Solitons and instantons (Amsterdam:
North-Holland), 1982.

\bibitem{CL}  N.H. Christ, T.D. Lee, Phys. Rev. {\bf D12}, 1606 (1975)

\bibitem{T}  E. Tomboulis, Phys. Rev. {\bf D12}, 1678 (1975)

\bibitem{J}  L. Jacobs, Phys. Rev. {\bf D13,} 2278 (1976)

\bibitem{GJeS}  J.-L. Gervais, A. Jevicki, B. Sakita, Phys. Rev. {\bf D12},
1038 (1975)

\bibitem{O}  K. Ohta, Phys. Rev. {\bf D43}, 2635 (1991)

\bibitem{O2}  K. Ohta, Nucl. Phys. {\bf A534}, 513 (1991)

\bibitem{HSaU}  A. Hayashi, S. Saito, M. Uehara, Phys. Lett. {\bf 246B}, 15
(1990)

\bibitem{UHSa}  M. Uehara, A. Hayashi, S. Saito, Nucl. Phys. {\bf A534}, 680
(1991)

\bibitem{SaOtY}  S. Saito, T. Otofuji, M. Yasumo, Prog. Theor. Phys. {\bf 75}%
, 68-83 (1986)

\bibitem{LiOk}  H. Liu, M. Oka, Phys. Rev. {\bf D40}, 883 (1989)

\bibitem{HSaU2}  A. Hayashi, S. Saito, M. Uehara, Phys. Rev. {\bf D43}, 1520
(1991)

\bibitem{HSaU3}  A. Hayashi, S. Saito, M. Uehara, Phys. Rev. {\bf D46}, 4856
(1992)

\bibitem{GoJa}  J. Goldstone, R. Jackiw, Phys. Rev. {\bf D11}, 1486 (1975)

\bibitem{Ts}  V.D. Tsukanov, (unpubl)

\bibitem{BD}  T.D. Bjorken, S.D. Drell, Relativistic quantum fields (Mc
Graw-Hill Book Company), 1965.
\end{thebibliography}
\end{document}